\documentclass[12pt,preprint]{aastex}

\def\Mstar{{{\rm M}\lower.5ex\hbox{$\star$}}}

\shorttitle{NEW PULSATING DBVs FROM SDSS}
\shortauthors{NITTA, KLEINMAN, KRZESINSKI ET AL.}

\begin{document}

\title{New Pulsating DB White Dwarf Stars from the Sloan Digital Sky Survey}

\author{A. Nitta\altaffilmark{1,2}, S.J. Kleinman\altaffilmark{1},
J.Krzesinski\altaffilmark{3}, S.O. Kepler\altaffilmark{4} T.S.
Metcalfe\altaffilmark{5},  Anjum S. Mukadam \altaffilmark{6}, Fergal
Mullally\altaffilmark{7}, R.E.  Nather\altaffilmark{8}, Denis J.
Sullivan\altaffilmark{9}, Susan E.  Thompson\altaffilmark{10}, D.E.
Winget\altaffilmark{8}}

\altaffiltext
{1}{Gemini Observatory, 670 N A'ohoku Pl.,Hilo, HI 96720 USA}
\altaffiltext
{2}{Subaru Telescope, 650 N A'ohoku Pl.,Hilo, HI 96720 USA}
\altaffiltext
{3}{Mount Suhora Observatory, Cracow Pedagogical University,
ul. Podchorazych 2, 30-084 Cracow, Poland }
\altaffiltext
{4}{Instituto de F\'{\i}sica, Universidade Federal do Rio Grande do Sul,
91501-970 Porto Alegre, RS Brasil}
\altaffiltext
{5}{High Altitude Observatory, NCAR, P.O. Box 3000, Boulder CO 80307 USA}
\altaffiltext
{6}{Department of Astronomy, University of Washington, Seattle,
WA 98195-1580 USA}
\altaffiltext
{7}{Department of Astrophysical Sciences, Princeton University, Princeton,
NJ 08544 USA}
\altaffiltext
{8}{Astronomy Department, University of Texas at Austin, Austin,
TX 78712, USA}
\altaffiltext
{9}{School of Chemical and Physical Sciences, Victoria University of
Wellington, New Zealand}
\altaffiltext
{10}{Dept.\ of Physics and Astronomy, University of Delaware, 223 Sharp
Laboratory, Newark, DE 19716 USA}

\begin{abstract}
We are searching for new He atmosphere white dwarf pulsators (DBVs)
based on the newly found white dwarf stars from the spectra obtained
by the Sloan Digital Sky Survey. DBVs pulsate at hotter temperature
ranges than their better known cousins, the H atmosphere white dwarf
pulsators (DAVs or ZZ Ceti stars). Since the evolution of white
dwarf stars is characterized by cooling, asteroseismological studies
of DBVs give us opportunities to study white dwarf structure at a
different evolutionary stage than the DAVs.  The hottest DBVs are
thought to have neutrino luminosities exceeding their photon
luminosities (Winget et al. 2004), a quantity measurable through
asteroseismology.  Therefore, they can also be used to study neutrino
physics in the stellar interior.  So far we have discovered nine
new DBVs, doubling the number of previously known DBVs. Here
we report the new pulsators' lightcurves and power spectra.
\end{abstract}

\keywords{stars:oscillations, stars:variables, general -- white dwarfs}

\section{Introduction}

White dwarf stars (WDs) are the endpoints of evolution for most
stars.  Their internal structures provide key clues into their
complex pre-WD evolution.  As WDs, their subsequent evolution is
dominated by cooling. The older they are, the cooler they become.
Why then, does there exist a range of temperatures within which we
hardly see any He atmosphere WDs (DBs) while we see both the H
atmosphere WDs (DAs) and non-DAs (He atmosphere DOs and DBs) at
both hotter and cooler temperature than this? This paradox is the
so-called ``DB gap'' (Fontaine \& Wesemael 1987).  Recently, Sloan
Digital Sky Survey (SDSS) data have shown us that the DB gap is not
completely void of DBs, but rather deficient in the number of DBs
(Eisenstein et al.  2006a).  The current best explanation for this
effect is based on WDs having specific layer masses (the large
gravity in a WD makes it compositionally stratified) which mix and
settle at certain temperatures, causing the surface ``flavor'' of
a WD to change with time and temperature (Fontaine \& Wesemael
1987). This explanation demands a thin H layer in at least a
substantial fraction of DAs.  However, there have been several works
(Fontaine et al. 1992; Clemens 1994; Fontaine et al. 1994; Robinson
et al. 1995; Kleinman et al.  1998; Benvenuto et al. 2002) suggesting
that perhaps all DAs have thick H layers and if so, spectral evolution
by the current model cannot happen.

Once a WD cools past the onset of its instability strip (at a
temperature primarily determined by its atmospheric composition and
total mass), it begins pulsating in a series of non-radial g-modes,
allowing us to study its interior via the technique of asteroseismology.
Asteroseismology, the study of stellar pulsations, is an important
way to directly measure quantities of the stellar interior. And
understanding the interior structure of the DBVs is one very important
way to address some of the mysteries of DB evolution.  Among the 9
DBVs known prior to our work, the first DBV discovered (Winget et
al. 1982), GD\,358, is by far the best studied WD pulsator. It has
had its internal structure substantially explored by asteroseismology
(Winget et al.  1994, Bradley \& Winget 1994; Vuille et al. 2000;
Metcalfe Salaris \& Winget 2002; Metcalfe 2003;  Kepler et al. 2005;
Metcalfe et al. 2005).  The results from the asteroseismological
investigations of GD\,358 (Winget et al., 1994) are impressive:
total mass of $0.61\pm0.03 M_{\sun}$, He layer mass of log$M_{He}/M_{\star}
= -5.7(+0.18, -0.30)$, $R_{\star}/R{\sun}=0.0127\pm0.0004$, He to
C transition zone thickness of about 8 pressure scale heights,
absolute luminosity log$L_{\star}/L{\sun}=-1.30 (+0.09, -0.12)$
hence a distance of $42\pm3pc$, weak magnetic field of $1300\pm300$G
and the measurements of radial differential rotation.  More recent detailed
model fitting techniques using genetic algorithms along with
improvements to the models have been successful in revealing even
more information.  We now have a measurement of the oxygen mass
fraction in the core which places constraints on both the nuclear
burning rate $^{12}C(\alpha, \gamma)^{16}O$ and even more detailed
structure information, such as the extent of the He/C envelope
beneath the pure He envelope (Metcalfe, Salaris \& Winget 2002;
Metcalfe 2003; Metcalfe et al. 2005).  Except for one other DBV,
the rest of the class have not been so forthcoming in revealing their
internal structures, primarily due to their lack of the abundance
of pulsation modes compared to GD\,358's over 100 detected frequencies.
CBS\,114 is a DBV which showed promise for successful asteroseismological
analysis by exhibiting a rich pulsation spectrum, but earlier
observational comparisons to the models produced a
$\mathrm{C}(\alpha,\gamma)\mathrm{O}$ nuclear burning
rate which was at odds with that obtained from GD\,358 (Handler,
Metcalfe \& Wood 2002). After several years of additional observations
of CBS\,114, which lead to identifying eleven independent pulsation
modes (four of which were new) along with improvements in pulsation
models and fitting techniques, Metcalfe et al.(2005) have achieved
new asteroseismological results for both stars which are now in
agreement with each other.  The one thing CBS\,114 did not show and
which GD\,358 did were the many fine structure splittings of the
pulsation modes caused predominantly by stellar rotation.  
Our understanding of the pulsation amplitude
determining mechanism on these stars is incomplete and we cannot
explain why we see significant fine-structure splitting in GD\,358
and not much in CBS\,114. We certainly do not believe it is due to
lack of rotation on CBS\,114's part though it could be due to the star
being observed near pole on. So the search goes on for a
third solvable pulsator to try and distinguish modes, models,
fits and reality in these objects.

Another important reason to study DBVs is that they are great cosmic
laboratories for high energy physics.  Winget et al. (2004) predict
that hot DBs should have significant plasmon neutrino production.
Their DB models suggest that 30,000K, $0.6M_{\sun}$ DBs have a
neutrino luminosity that is 1.8 times higher than their photon
luminosity. On the cool end, 22,000K,  $0.6M_{\sun}$ DBV models
have a neutrino luminosity less than half of their photon luminosity.
Thus the hottest DBVs should be losing energy and cooling significantly
faster than the cooler ones.  Since a pulsation mode's period is a
function of temperature, we can directly measure a star's cooling
rate by measuring a mode's rate of period change (e.g. Kepler et
al.  2005b). And thus, the DBVs may be quite revealing laboratories
for neutrino physics.

Finally, an increase in the number of known DBVs will help us
understand their properties as a group.  Clemens (1994) and Kleinman
(1995, 1998) found that the DA pulsators break down nicely into two
distinct classes, each subclass exhibiting common class properties
which they have used to investigate the dynamics of the pulsation
mechanism in these stars.  By increasing the number of known DBVs,
we can search for possible subclass distinctions.  Nather, Robinson
\& Stover (1981) noted that the interacting binary white dwarf stars
will each eventually form a single DB at the end of their evolution.
This means that there may be more than one evolutionary channel
leading to the DBs.  Perhaps we will find two distinct classes,
each of them retaining the evidence of their evolutionary paths in
their pulsation structures.

SDSS is a photometric and spectroscopic survey of the sky covering
about 10,000 square degrees around the Northern Galactic
cap (York et al. 1996; Stoughton et al. 2002; Gunn et al. 1998;
Gunn et al.  2000). In SDSS's Sixth Data Release  (Adelman-McCarthy,
et al. 2008), there are photometry of close to 10,000 square degrees
in five filters (Fukugita et al. 1996) and 1.27 million spectra.
Although the survey's main goal was to produce a 3D map of the large
scale structure of the universe, it also contains data on many galactic
stellar objects, including WDs.  SDSS data provide the perfect basis
set for finding new DBVs which will eventually help solve the DB
Gap mystery, measure the neutrino production rates inside the DBs,
as well as  answer some other questions about WD structure and
evolution.  Kleinman et al. (2004) published the first WD catalogue
based on the spectra obtained by SDSS.  and doubled the number of
then known WDs. The newest WD catalogue from the SDSS (Eisenstein
et al.  2006b, DR4 WD catalogue hereafter) has almost quadrupled
the number of WDs. Among the new WDs are DBs whose physical parameters
determined from model fitting suggest they are inside the instability
strip.  Therefore, we started a project to search for new DBVs using
our spectroscopic fits to SDSS spectra, originally from Kleinman
et al.  (2004) and later using the DR4 WD catalogue, to identify
likely DBV candidates and follow them up with time-series photometry.
This survey is the counterpart to the search for new SDSS DAVs
reported by Mukadam et al. (2004), Mullally et al.(2005), Kepler
et al.  (2005a) and Castanheira et al. (2006a, 2007).

\section{Observations}

We selected our DBV candidates based on  the effective temperatures
published in the SDSS WD catalogues (Kleinman et al. 2004; Eisenstein
et al. 2006).  As described in those works, each spectrum was fit
with Detlev Koester's atmosphere models (Koester et al.  2001) to
obtain an effective temperature and surface gravity. The DB models
used in the catalogues are pure He models. Beauchamp et al. (1994) showed
the physical parameters of the model fit of DBs can change if He atmosphere
models with trace amount of H are used.  Since we do not know how
much H, if any, our candidate SDSS DBs have, the pure He atmosphere
models fits are as good as any other. Given the currently known coolest
DBV being 21,800K (Beauchamp et al.  1994; Castanheira et al. 2006b),
we chose to select all DBs with effective temperatures higher than
21000K as DBV candidates.  The blue edge of the instability strip
is currently defined by EC\,20058, the second hottest DB known
(Beauchamp et al.  1999; Sullivan et al.  2008) prior to the new
DBs discovered by the SDSS. The hottest DB known prior to the SDSS
is PG0112+104 with $T_{\rm eff}=31,500$K which defines the cool end
of the DB gap. Time series observations of this star have not
detected any pulsations (Provencal 2006).  Nonetheless, given a
boundary determined only by one object, we decided to place no upper
limit on our candidate stars' effective temperatures.

We observed our DBV candidates using the Argos CCD camera (Nather
\& Mukadam 2004) on the 2.1m telescope at McDonald Observatory,
SPICam on the 3.5m telescope at Apache Point Observatory and SOAR
Optical Imager (SOI) on the Southern Astrophysical Research Telescope
(SOAR).  More than half of the new H atmosphere white dwarf variables
(DAVs) reported in the past few years have been discovered using
Argos (Mukadam et al. 2004; Mullally et al.  2005; Castanheira et
al.  2006a).  We observed and reduced the data from Argos in the
same manner as described in Mukadam et al. (2004) and Mullally et
al.  (2005). Exposure times ranged from 5s to 30s, depending on the
brightness of the target and condition of the sky.  The readout
time was negligible due to the use of a frame transfer detector.
For some of the objects, we used a BG40 filter to suppress the
redder portion of the flux which is dominated by noise. After we
applied bias and flat field corrections to all CCD frames, we
extracted sky-subtracted lightcurves via aperture photometry for
the variable candidates and at least one comparison star in the
field.  We then divided the target star's lightcurve by the sum of
the comparison stars' lightcurves to take out any transparency
variations in the sky.  We normalized the result so that the average
brightness of the star is equal to 0 and the lightcurve shows the
fractional intensity variation, and applied a barycentric correction
to the times.  The resulting lightcurves for the new
DBVs are shown in the left panel of Figure~1.

SPICam was not built for fast time series data acquisition and
therefore we binned and used partial readout to achieve a reasonable
duty cycle for this project. The binning and window size of the
chip depended on the seeing and field of the target since we needed
at least one comparison star. Once we acquired the data, we followed
a similar procedure as with Argos data to produce our lightcurves.

We used SOI to discover our 9th DBV. SOI has also contributed to
discoveries of 18 new DAVs (Kepler et al. 2005; Castanheira et al.
2006).  It is a CCD camera with reasonably fast readout time (6.3s).
We used 30s integration time for the data we gathered on
SDSS~J085202.44+213036.5 Again, we followed a similar procedure as
with Argos data to produce our lightcurves.

Table~1 is our journal of observations.  We tried to observe each
object for at least two~hours on two separate occasions. The second
observation is to confirm and test the results of the first
observation.  As you can see from Table~1, we have been able to get
the second observation for five of the new DBVs, but not for all
of the objects reported in this paper.  For the DBs which did not
show pulsations during the first observations, additional data are
still very important.  The lack of variability in the first observation
may simply be due to amplitude modulations or beating of closely
spaced modes which are not resolved in our $\sim$2 hours observations.
It is also important to obtain a good amplitude limit (1mma or
smaller) to which we see no variability since some currently known
pulsators have similarly small amplitudes.  We note that some of
the DAs which had no detectable pulsations in Mukadam et al. (2005)
turned out to be DAVs after additional observations lowered the
detectable amplitude limit (Castanheira, et al. 2006).  Both these
examples suggest more data are still needed for many of our DBs
which did not show pulsations.  

\section{New Pulsating DB White Dwarf Stars}

Figure~1 shows the lightcurves and their Fourier transforms for the
nine new DBVs we have found so far.  We list the frequencies,
periods and the amplitudes of the large observed peaks in the FTs
in Table~2.

The g magnitudes from the SDSS imaging data, the plate, MJD and
fiber number which specify unique spectra used for the model fitting,
the effective temperature, surface gravity and their uncertainties
of each observed object are given in Table~3. The last column in
Table~3 indicates if the object was found to vary. If we saw
no variability, then this column contains the amplitude limit (in
mma) we currently have. The amplitude limit is defined as three
times the average noise between 1000-10,000$\mu$Hz. For equally
spaced data, this limit translates into a $0.1\%$ probability of
identifying false peak as a real one (e.g. Kepler 1993).  This frequency
range corresponds to periods of 100$s$ to 1000$s$ where the pulsations
in DBVs have been detected. We also note that some of the lightcurves
contain noise at low frequencies (less than few hundred $\mu$Hz
which corresponds to several thousand seconds and longer in period),
probably due to transparency variations or thin cirrus. If we
included this noise in our estimate, our amplitude limits would
have been higher and not reflective of our true ability to detect
variation within the frequency range of interest.

In Figure~2, we plot the effective temperatures and the surface
gravities for DBs in the DR4 WD catalogue.  Newly found DBVs,
represented by large solid dots with their uncertainties in effective
temperature and surface gravity, cluster around $T_{\rm eff} \sim
25,000$K, although many more objects still need observation (the hollow
dots). We did not plot each set of error bars to avoid clutter in the
figure.  Many of the DBs for which we did not see any variability
(represented by squares in Figure~2) have not been observed a second
time, mainly because we have not yet had time to do so. As you can see
from Table~1, only two objects (SDSS J090409.03+012740.9 and SDSS
J141258.17+045602.2) were observed more than once with combined
amplitude limits of 3.5\,mma and 2.6\,mma,
respectively.  These amplitude limits are by no means good enough to
call them non-pulsators since some WD pulsators are known to have lower
amplitudes than this.  Our current results are consistent with, but do
not demand, a pure DBV instability strip.  We need to eventually
achieve at least 1\,mma detection limit for all the DBV candidates we
observe before investigating the purity of the instability strip.

We observed four DBs with $T_{\rm eff}>30,000$K, i.e. DBs in the
``DB gap'', but we did not see any pulsations so far. Like other
DBs we observed and not detected pulsations, these objects need
to be followed up before they can be declared non-pulsators.  In
the past, the instability strip was defined by the 9 known DBVs
shown by triangles in Figure~2.  The blue edge of the instability
strip was defined by one DBV, EC20058.  We have not found any
pulsator hotter than EC20058 and hence the best chance of determining
the neutrino production rates still lies with this star.

\section{Summary}

From the DR4 WD catalogue, we have about 70 DBV candidates brighter
than  $g=20$mag.  To date, we have observed 29 of them and found
nine new DBVs, doubling the number of known DBVs.  We seek an increased number
of DBVs to help us understand their group properties, better determine
the location of the instability strip, and perhaps find hot DBVs
we can use to measure their cooling rates and place a limit on the
neutrino production rate in their interiors. Based on these statistics,
we can expect at least another 12 new DBVs from the DR4 sample and
20 more from DR6. These are probably lower limits though,  since
we suspect additional observations of our 29 currently observed
objects will probably reveal new low amplitude pulsators as well.

\acknowledgments

Funding for the SDSS and SDSS-II has been provided by the Alfred P.
Sloan Foundation, the Participating Institutions, the National
Science Foundation, the U.S. Department of Energy, the National
Aeronautics and Space Administration, the Japanese Monbukagakusho,
the Max Planck Society, and the Higher Education Funding Council
for England. The SDSS Web Site is http://www.sdss.org/.

The SDSS is managed by the Astrophysical Research Consortium for
the Participating Institutions. The Participating Institutions are
the American Museum of Natural History, Astrophysical Institute
Potsdam, University of Basel, Cambridge University, Case Western
Reserve University, University of Chicago, Drexel University,
Fermilab, the Institute for Advanced Study, the Japan Participation
Group, Johns Hopkins University, the Joint Institute for Nuclear
Astrophysics, the Kavli Institute for Particle Astrophysics and
Cosmology, the Korean Scientist Group, the Chinese Academy of
Sciences (LAMOST), Los Alamos National Laboratory, the
Max-Planck-Institute for Astronomy (MPIA), the Max-Planck-Institute
for Astrophysics (MPA), New Mexico State University, Ohio State
University, University of Pittsburgh, University of Portsmouth,
Princeton University, the United States Naval Observatory, and the
University of Washington. 

This work was partially supported by the National Science Foundation
through an Astronomy \& Astrophysics Postdoctoral Fellowship (to
T.S.M.) under award AST-0401441.



\clearpage

\begin{figure}
\plottwo{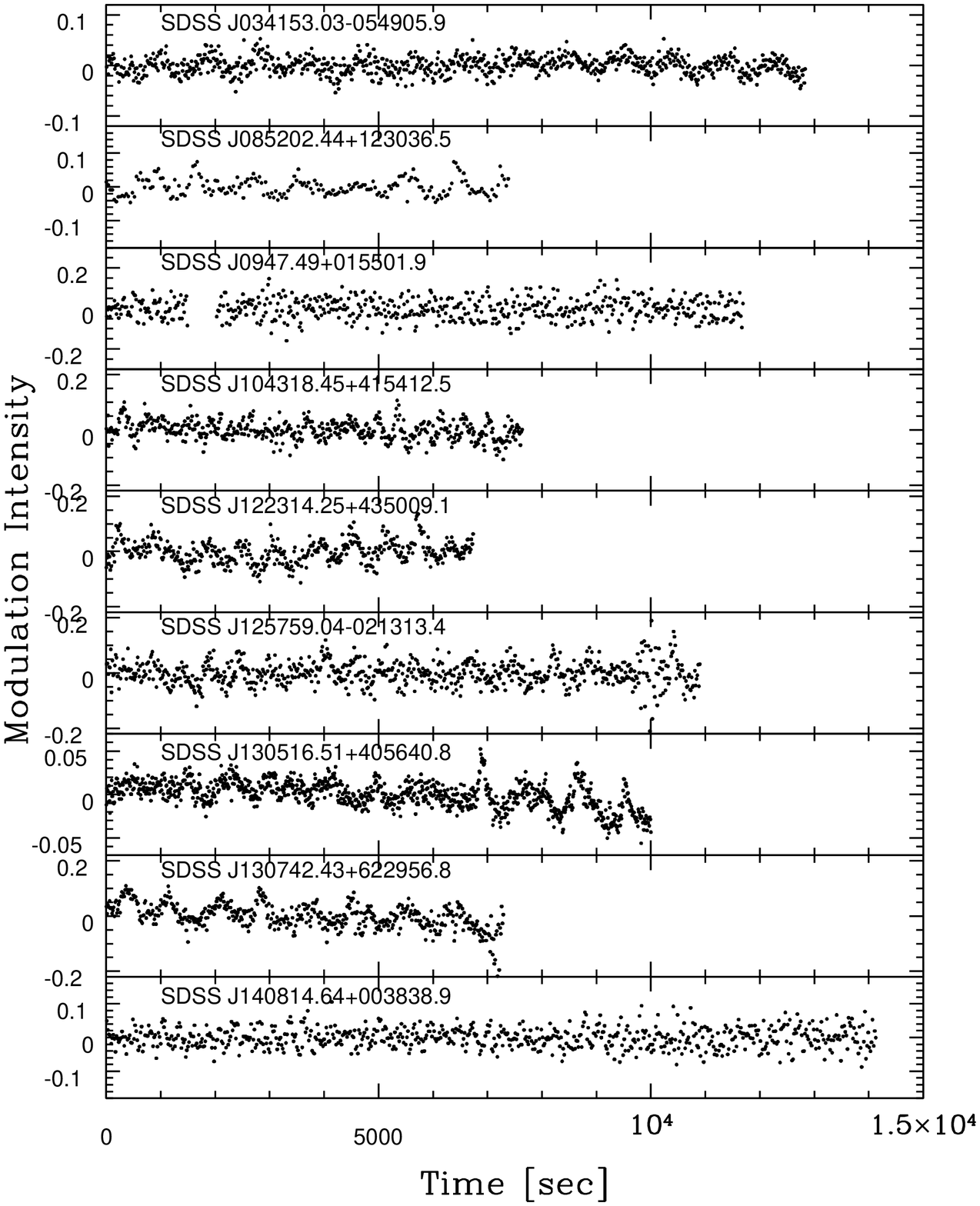}{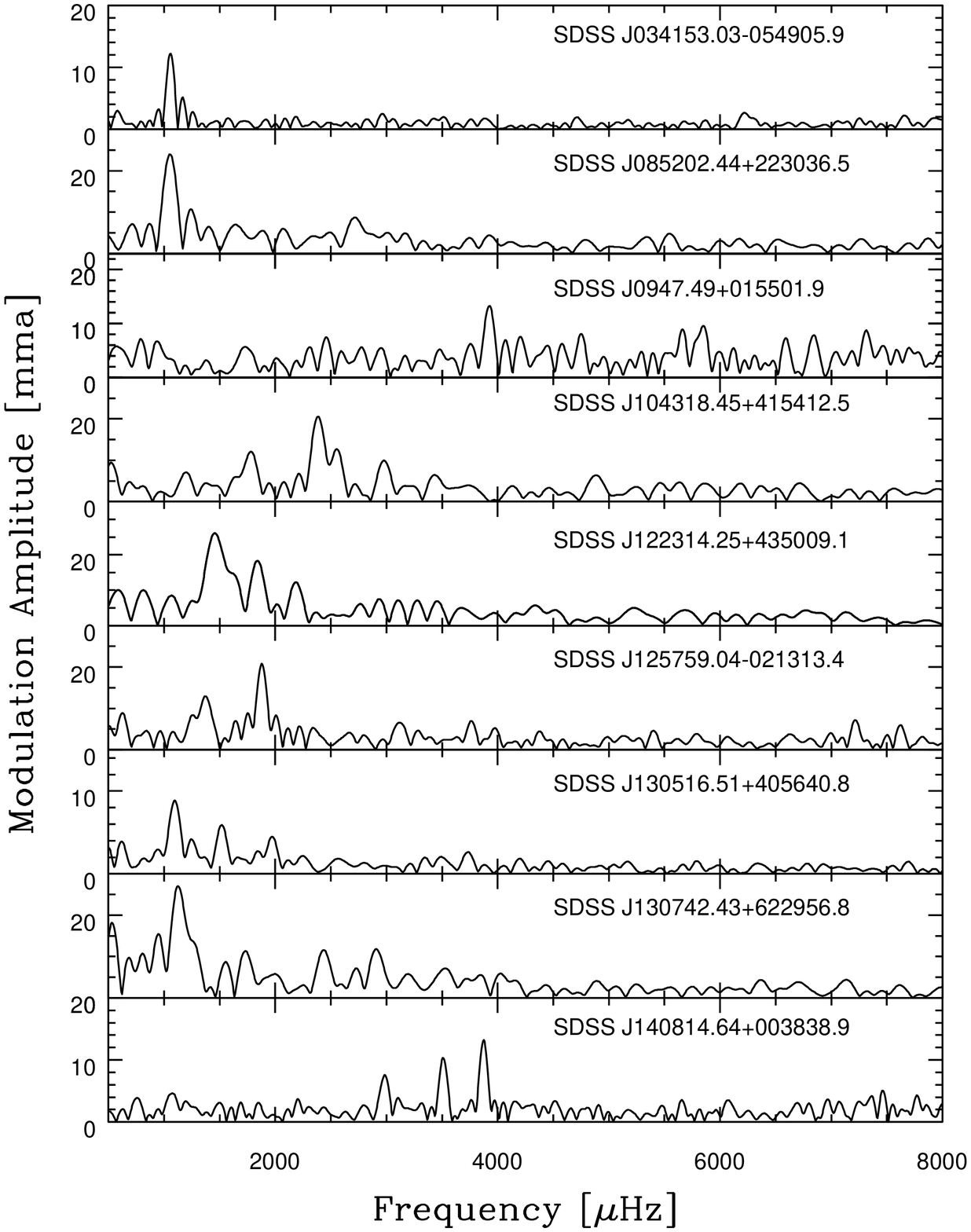}
\label{lcft}
\caption{Lightcurve (the left panel), Fourier transform (the right
panel) 
of the 9 new DBVs reported in this paper. The lightcurve of
SDSS~J140814.64+003838.9 was binned by two (i.e. changing the sampling
rate from 10s to 20s) to show the pulsation better in the figure, but
the FT was calculated from the unbinned data. SDSS~J0947.49+015501.9's FT,
perhaps, is not as visually convincing as other new DBVs shown here.
The FT of the second observation of this target also shows the largest peak at a
consistent frequency as the data shown here with similar significance.
} 
\end{figure}

\clearpage

\begin{figure}
\plotfiddle{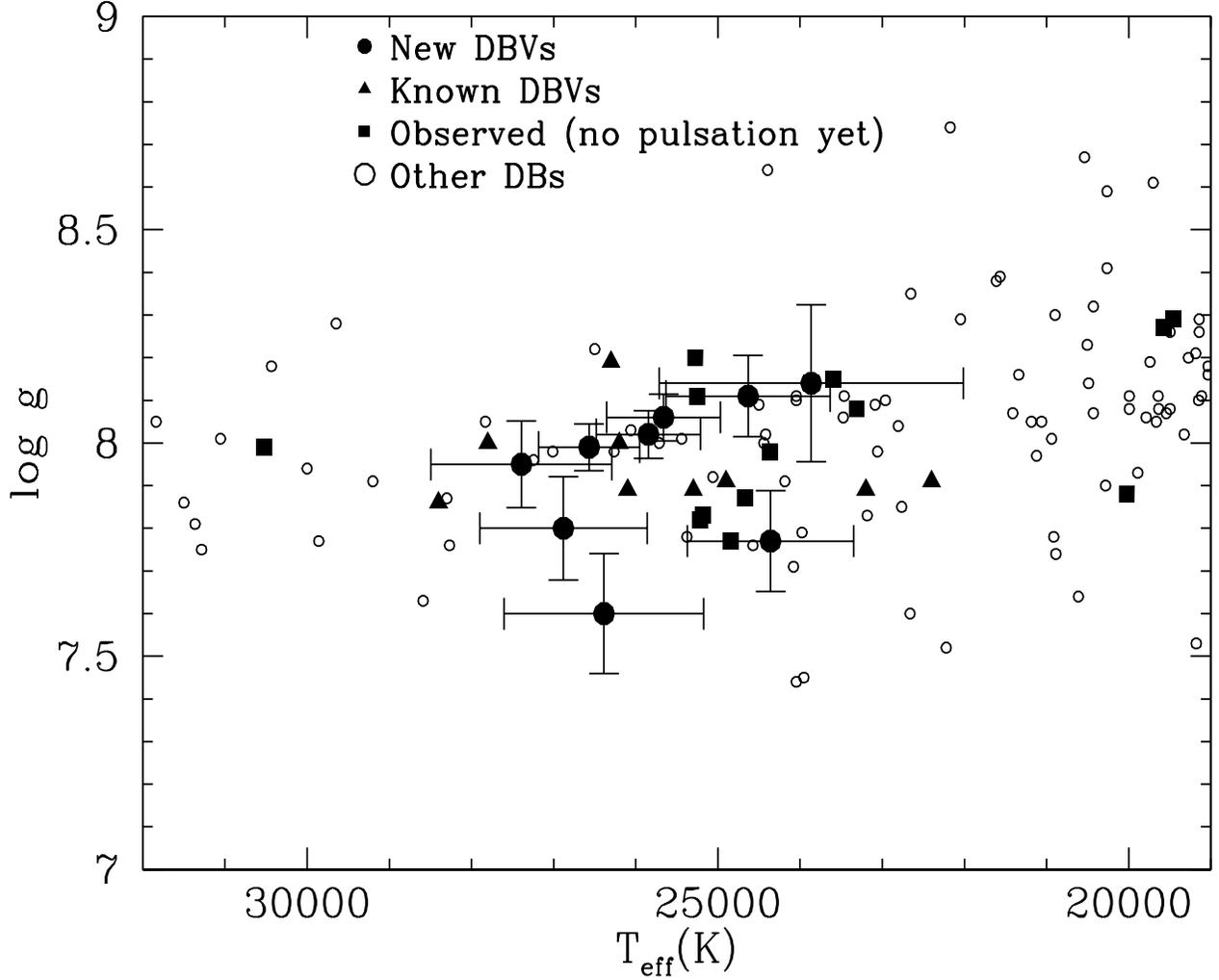}{0in}{270}{400}{500}{0}{0}
\label{tg}
\caption{Here we indicate the T${\rm eff}$ and the ${\rm log}$~g of the
DBVs we found by black dots, the previously known DBVs by triangles,
the observed DBs by squares and all other DBs in the SDSS DR4 by hollow
dots. We plot only the error bars of the new DBVs to avoid clutter in
the diagram. The previously known DBVs' physical parameters were taken
from Beauchamp et al. (1994). We only quote results from their pure He
atmosphere model fits since we use pure He atmosphere models for all DBs
from the SDSS.  Their
models and spectral fitting techniques and ours are different.
Therefore there are probably some offset/differences in the temperature
and gravity scale compared to those from ours. Some of the observed
DBs are outside the temperature range shown here. 
} 
\label{instability}
\end{figure}

\clearpage
\begin{table}
\begin{center}
\footnotesize
\caption{\small Journal of Observations. The new DBVs reported in this paper are
marked with an * next to the object name.}
\begin{tabular}{l|c|r|c|c}
\tableline
\tableline
Name & Date & Length & Run number\tablenotemark{1}&Filter \\
SDSS J & & [s] & &\\
\tableline

001529.75+010521.4 & 2003 Nov 27 & 12103.06 & APO & none  \\
031609.12-062556.8 & 2007 Feb 13 & 9000.00 & A1446 & BG40 \\
034153.03--054905.9* & 2003 Dec 03 & 12832.5 & A0797 & none\\
                   & 2003 Dec 26 &  4380.0  &  A0811 & BG40\\
                   & 2003 Nov 27 & 7575.13 & APO  &none\\
081904.19+354255.8 & 2007 Feb 13 & 12380.0 & A1447 & BG40\\
085202.44+213036.5* & 2008 Mar 15 & 7381.15 & SOAR & B \\
                    & 2008 May 06 &  7038.50 & SOAR & B \\
085950.30--000339.6	& 2003 Dec 27 & 10650.0 & A0818 & none \\
090409.04+012741.0	& 2003 Dec 26 &  7085.0 & A0813 & none \\
                    & 2003 Nov 27 & 9026.21 & APO & none \\
090456.13+525029.9 & 2003 Mar 10 & 9026.21 & APO & none \\
092200.98+000834.4	& 2003 Dec 24 &  6337.5 & A0804 & none  \\
094749.40+015501.9* & 2003 Dec 22 &  8692.5 & A0799 & none \\
                   & 2003 Dec 30 & 11690.0 & A0828  & none\\
095256.69+015407.7	& 2003 Dec 23 &  5505.0 & A0801 &none  \\
095455.11+440330.3 & 2007 Feb 15 & 14055.0 & A1451 & BG40 \\
095649.55+010812.4 & 2003 Apr 27 & 11965.58 & APO & none  \\
101131.88+050729.3 & 2005 Apr 01 & 7000.0 &  A1022 & none \\
101502.95+464835.3 & 2005 Apr 01 & 6000.0 & A1017 & none \\
104318.45+415412.5* & 2005 Apr 05 &  7635.0 & A1027 & none \\
105929.60+554039.2 & 2005 Apr 01 & 5360.0 & A1018 & none \\
122241.28--003614.4	& 2003 Dec 26 &  6547.5 & A0815 &none  \\
122314.25+435009.1* & 2005 Apr 05 &  6735.0 & A1028 & none \\
                   & 2007 Feb 13 & 7240.0 & A1448 & BG40\\
125759.04--021313.4* & 2003 Apr 01 & 10897.5 & A0626 & BG40 \\
                   & 2003 Apr 27 &  4968.04 & APO & none\\
130516.51+405640.8* & 2005 Apr 02 & 10000.0 & A1019 & none \\
130742.43+622956.8* & 2005 Apr 03 &  7290.0 & A1023 & none \\
131148.49+053847.6 & 2007 Feb 13 & 7080.0 & A1449 & BG40 \\
133215.95+640656.3 & 2003 May 27 & 10576.08 & APO & none  \\
135610.31--002230.6	& 2003 Dec 30 &  4822.5 & A0829 &none  \\
140814.64+003838.9* & 2003 Mar 31 & 14145.0 & A0602  & none\\
                   & 2004 Apr 20 &  6930.0 & A0868 & BG40 \\
                   & 2003 Mar 24 & 17084.99 & APO & none\\
141258.17+045602.2 & 2003 Apr 30 & 4195.0 & A0623 & BG40\\
                   & 2003 Mar 10 & 5126.25 & APO & none  \\
                   & 2003 May 27 & 6694.15 & APO & none  \\ 
                   & 2003 Apr 27 & 9164.86 & APO & none  \\
231324.25--001636.8	& 2003 Dec 26 &  4552.5 & A0810 & none  \\
                    & 2003 Dec 30 &  5655.0 & A0825 & none  \\
235322.16+002653.9 	& 2003 Dec 22 &  7027.5 & A0796 & none  \\

\tableline
\end{tabular}
\tablenotetext{1}{Texas data have run numbers starting with a letter A
followed by a 4 digit number. APO data do not have a run number and are
indicated by ``APO'' in this column.}

\end{center}
\end{table}

\clearpage
\begin{table}
\caption{\small Observed periods and amplitudes in the new DBVs. We do not
currently  have the
resolution to detect any multiplets or closely spaced modes.}
\begin{center}
\footnotesize
\begin{tabular}{lccr}
\hline
Object &
Frequency  &
Period  &
Amplitude  \\
SDSS J &
[$\mu$Hz] &
[s] &
[mma] \\

\hline
034153.03-054905.8 & 1060.5 &  942.0 &   12.2  \\
085202.44+213036.5 & 1051.9 &  950.7  & 20.8 \\
094749.40+015501.8 & 3923.9 &  254.9 &   13.3 \\
104318.45+415412.5 & 2382.6 & 419.7 &  20.6  \\
122314.25+435009.1 & 1456.4 & 686.6 &  26.1   \\
 &1838.2 &  544.0 &   18.3 \\
125759.03-021313.3 &  1371.6 & 729.1 &  13.0  \\
 &  1880.6 &  531.7 &  20.8 \\
130516.51+405640.8&  1095.9 & 912.5  &  8.9  \\
 &  1520.1 & 657.9  &  5.9 \\
130742.43+622956.8 & 1124.1 & 889.6 &  27.0  \\
140814.63+003838.9 & 2983.5 & 335.2 &   7.6   \\
 &  3506.7 & 285.2 &  10.3 \\
 &  3874.4 & 258.1 &  13.2 \\
\tableline
\end{tabular}
\end{center}
\end{table}

\clearpage
\begin{table}
\caption{\small SDSS data on all observed DBs. The top section of the table
details the objects that showed variability during at least one
observation. Separated by a double horizontal line, the second half
of the table lists the objects for which we have not (yet?) seen
significant variability.  In the status section, we note new variable
objects by ``DBV''. For objects in which we have not detected
variability, we give the amplitude limit in mma in the status
section.  If we have only observed an object once, then we add a
``(1)''.  Due to lack of observing time and a large number of
candidates, we have yet been able to observe all DBV candidate
objects, nor all these a second time. The physical parameters here
come from fitting SDSS DR6 spectral data with a denser, but
otherwise consistent,  model grid than used in the DR4 WD catalog.}
\footnotesize
\label{abun}
\begin{center}
\begin{tabular}{lrrrrrrrrl}
\tableline
Object  &
Plate &
Fiber &
MJD &
g &
T${\rm eff}$ & 
$\sigma_{\rm Teff}$ &
${\rm log g}$ &
$\sigma_{\rm logg}$ &
Status \\
SDSS J & & & & [mag] & [K] & [K] & & & \\
\tableline
\tableline
034153.03-054905.8 & 462 & 506 & 51909 & 18.25 & 25087 & 524 & 8.02 & 0.062 & DBV\\
085202.44+213036.5 & 2280& 604& 53680 & 18.50& 25846 & 6361 & 8.02 & 0.056 & DBV\\
094749.40+015501.8 & 480 & 520 & 51989 & 19.95 & 23453 & 1659 & 8.13 & 0.192 & DBV\\
104318.45+415412.5 & 1361 & 155 & 53047 & 18.95 & 26291 & 919 & 7.77 & 0.138 & DBV(1)\\
122314.25+435009.1 & 1371 & 205 & 52821 & 18.98 & 23442 & 1069 & 7.84 & 0.127 & DBV\\
125759.03-021313.3 & 338 & 436 & 51694 & 19.16 & 25820 & 1296 & 7.57 & 0.151 & DBV\\
130516.51+405640.8 & 1458 & 21 & 53119 & 17.46 & 24080 & 414 & 8.14 & 0.056 & DBV(1)\\
130742.43+622956.8 & 783 & 513 & 52325 & 18.83 & 23841 & 913 & 8.14 & 0.097 & DBV(1)\\
140814.63+003838.9 & 302 & 490 & 51688 & 19.19 & 26073 & 1227 & 7.98 & 0.117 & DBV\\
\tableline
\tableline
001529.74+010521.3 & 389 & 530 & 51795 & 18.94 & 34379 & 1079 & 7.96 & 0.163 & 8.20(1)\\
031609.12-062556.8 & 459 & 605 & 51924 & 19.97 & 24478 & 2520 & 7.96 & 0.222 & 17.0(1)\\
081904.19+354255.8 & 826 & 422 & 52295 & 18.22 & 22540 & 867 & 8.18 & 0.079 & 4.80(1)\\
085950.29-000339.6 & 469 & 49 & 51913 & 20.19 & 23729 & 2391 & 8.12 & 0.291 & 13.3(1)\\
090409.03+012740.9 & 470 & 442 & 51929 & 17.96 & 23183 & 533 & 7.95 & 0.062 & 4.28\\
090456.11+525029.8 & 552 & 547 & 51992 & 18.95 & 37584 & 953 & 7.99 & 0.091 & 10.1(1)\\
092200.97+000834.3 & 474 & 388 & 52000 & 18.56 & 22581 & 769 & 8.10 & 0.074 & 7.56(1)\\
095256.68+015407.6 & 481 & 513 & 51908 & 17.50 & 32920 & 323 & 8.16 & 0.041 & 4.84(1)\\
095455.11+440330.3 & 942 & 275 & 52703 & 18.18 & 20072 & 368 & 8.29 & 0.064 & 5.85(1)\\
095649.55+010812.4 & 481 & 20 & 51908 & 20.48 & 17125 & 1257 & 7.37 & 0.261 & 13.0(1)\\
101131.88+050729.3 & 574 & 331 & 52355 & 18.97 & 24301 & 984 & 7.71 & 0.115 & 8.98(1)\\
101502.95+464835.3 & 944 & 328 & 52614 & 18.61 & 23312 & 830 & 8.01 & 0.076 & 7.24(1)\\
105929.60+554039.2 & 908 & 317 & 52373 & 18.47 & 24742 & 571 & 8.17 & 0.101 & 8.46(1)\\
122241.27-003614.4 & 288 & 63 & 52000 & 18.10 & 24023 & 676 & 8.21 & 0.073 & 4.66(1)\\
131148.49+053847.6 & 850 & 522 & 52338 & 17.65 & 20249 & 268 & 8.30 & 0.041 & 11.7(1)\\
133215.93+640656.2 & 603 & 118 & 52056 & 18.41 & 21365 & 1694 & 7.99 & 0.097 & 9.73(1)\\
135610.32-002230.6 & 301 & 232 & 51641 & 19.38 & 18584 & 397 & 8.20 & 0.149 & 13.1(1)\\
141258.17+045602.2 & 583 & 432 & 52055 & 17.35 & 30343 & 329 & 7.97 & 0.038 & 2.88\\
231324.24-001636.9 & 381 & 72 & 51811 & 19.83 & 19588 & 1987 & 7.93 & 0.298 & 3.19\\
235322.16+002653.8 & 386 & 549 & 51788 & 19.71 & 25012 & 1800 & 8.15 & 0.203 & 13.0(1)\\
\tableline
\end{tabular}
\end{center}
\end{table}


\begin{thebibliography}{}


\bibitem {dr6} Adelman-McCarthy, J., et al. 2008, ApJS, in press

\bibitem {beauchamp} Beauchamp, A., Wesemael, F., Bergeron, P., Fontaine,
G., Saffer, R. A., Liebert, J., Brassard, P.  1999, ApJ, 516, 887

\bibitem {argentina} Benvenuto, O. G.; Corsico, A. H.; Althaus, L.
G.; Serenelli, A. M. 2002, MNRAS, 322, 299


\bibitem {bradleywinget} Bradley, P.A. \& Winget, D.E. 1994, 430, 850

\bibitem {barbara} Castanheira, B. G, et al.
2006a, A\&A, 450, 227

\bibitem {BarbaraIAU} Castanheira, B. G., Kepler, S. O., Handler, G.,
Koester, D. 2006b, A\&A, 450, 331

\bibitem {cleanup} Castanheira, B.G., et al.
2007. A\&A, 462, 989

\bibitem {christhesis} Clemens, J.C., 1994, PhD Thesis, University of Texas at Austin


\bibitem {dbgap} Eisenstein, Daniel J., et al.
2006a, AJ, 132, 676

\bibitem {dr4cat} Eisenstein, Daniel J., et al.
2006b, ApJS, 167, 40

ApJ, 488, 375

\bibitem{fw} Fontaine, G. \& Wesemael, F. 1987, in IAU Colloq. 95,
2nd Conf. on Faint Blue Stars, ed. A. G. D. Phillip, D. S. Hayes, \&
J. Liebert (Schenectady,Davis), 319

\bibitem{g226} Fontaine, G., Brassard, P., Bergeron, P. \& Wesemael, F.
1992, ApJ, 399, L91

\bibitem{thicklayer} Fontaine, G., Brassard, P., Wesemael, W. \& Tassoul,
M. 1994, ApJ, 428, L61

\bibitem {sdssfilters} Fukugita, M., Ichikawa, T., Gunn, J.E., Doi,
M., Shimasaku, K. \& Schneider, D.P. 1996, AJ, 111, 1748

\bibitem {camera} Gunn, J.E., et al. 1998, AJ, 116, 3040 

\bibitem {telescope} Gunn, J.E., et al. 2006, AJ, 131, 2332


\bibitem {gerald} Handler, G., Metcalfe, T.S. \& Wood, M.A. 2002, MNRAS, 335, 698

\bibitem {models} Koester, D., et al. 2001, A\&A, 378, 556

\bibitem {review} Kepler, S.O. 1993, Baltic Astron, 2, 515

\bibitem {kepler} Kepler, S.O., et al.
2005a, A\&A, 442, 529

\bibitem {g117} Kepler, S.O., et al.
2005b, ApJ, 634, 1311

\bibitem {scotthesis} Kleinman, S.J., 1995, PhD Thesis, University of Texas at Austin

\bibitem {cooldav} Kleinman, S.J., et al.
1998, ApJ, 495, 424

\bibitem {wdcat} Kleinman, S.J., et al.
2004, ApJ, 607, 426

\bibitem {dq} Liebert, J.,  et al.
2004 AJ, 126, 2521

\bibitem[Metcalfe et al.(2002)]{2002ApJ...573..803M} Metcalfe, T.~S.,
Salaris, M., \& Winget, D.~E.\ 2002, \apj, 573, 803

\bibitem[Metcalfe(2003)]{2003ApJ...587L..43M} Metcalfe, T.~S.\ 2003,
\apjl, 587, L43 


\bibitem[Metcalfe et al.(2005)]{2005A&A...435..649M} Metcalfe, T.~S.,
Nather, R.~E., Watson, T.~K., Kim, S.-L., Park, B.-G., \& Handler, G.\
2005, \aap, 435, 649

\bibitem {anjum} Mukadam, Anjum S., et al.
2004, ApJ, 607, 982

\bibitem {fergal} Mullally, F.,  et al.
2005, ApJ, 625, 966

\bibitem {g61-29} Nather, R.E., Robinson, E.L. \& Stover, R.J. 1981, ApJ,
244, 269

\bibitem {argos} Nather, R.E., \& Mukadam, Anjum, S. 2004, ApJ, 605, 849


\bibitem {judi} Provencal, J.L. 2006, private communication

\bibitem{rob} Robinson, E.L., et al. 1995, ApJ, 438, 908

\bibitem {edr} Stoughton, C., et al. 2002, AJ, 123, 485 

\bibitem {denis} Sullivan, D., et al.  2008, MNRAS, 387, 137

\bibitem {evil} Vuille, F.,  et al.
2000, MNRAS, 314, 689

\bibitem {gd358discovery} Winget, D.E., Robinson, E.L., Nather, R.E., \&
Fontaine, G. 1982, \apj, 262, L11

\bibitem {gd358} Winget, D.E. et al.
1994, ApJ, 430, 839

\bibitem {plasma} Winget, D.E., Sullivan, D. J., Metcalfe, T. S., Kawaler,
S. D., Montgomery, M. H 2004, ApJ, 602, 109

\bibitem {york} York, D.G., et al. 2000, AJ, 120, 1579

\end{thebibliography}
\end{document}